# Exploration of metastable A-site-ordered perovskites (Ca,Ba)FeO$_{3-\delta}$ by computationally-guided multi-step synthesis

Masaho Onose[1,2], Hidefumi Takahashi[1,3], Hajime Sagayama[4], Yuichi Yamasaki[5], and Shintaro Ishiwata[1,3,*]

[1] *Division of Materials Physics, Graduate School of Engineering Science, Osaka University, Toyonaka, Osaka 560-8531, Japan*
[2] *Department of Applied Physics, the University of Tokyo, Bunkyo-ku, Tokyo 113-8656, Japan*
[3] *Center for Spintronics Research Network (CSRN), Graduate School of Engineering Science, Osaka University, Toyonaka, Osaka 560-8531, Japan*
[4] *Institute of Materials Structure Science (IMSS), High Energy Accelerator Research Organization (KEK), Tsukuba, Ibaraki 305-0801, Japan*
[5] *National Institute for Materials Science (NIMS), Tsukuba, Ibaraki 305-0047, Japan*

**Abstract**

Perovskite-type iron oxides with Fe$^{4+}$ ions have attracted much attention for their versatile helimagnetic phases. While the introduction of a layered A-site ordered structure to AFeO$_3$ with Fe$^{4+}$ ions potentially lead to novel helimagnetic phases, the synthetic pathway spanning high pressure range is apparently difficult to elucidate. Here, we explored new A-site ordered perovskite-type iron oxides (Ca,Ba)FeO$_{3-\delta}$ with Fe$^{4+}$ ions with the support of first-principles calculations evaluating thermodynamic stability at selected pressures and chemical compositions. Among the six types of putative A-site ordered perovskites with and without oxygen vacancy, only two types of oxygen-deficient perovskites CaBaFe$_2$O$_{6-\delta}$ and Ca(Ba$_{0.9}$Ca$_{0.1}$)$_2$Fe$_3$O$_{9-\delta}$ ($\delta\sim1$) were successfully obtained by high-pressure synthesis, being consistent with the DFT-based convex-hull calculations. Considering the evaluated stability of the putative perovskites at selected pressures, we adopted low-temperature topotactic oxidation using ozone at ambient pressure and obtained the oxidized perovskites CaBaFe$_2$O$_{6-\delta}$ ($\delta\sim0.4$) and Ca(Ba$_{0.9}$Ca$_{0.1}$)$_2$Fe$_3$O$_{9-\delta}$ ($\delta\sim0.6$), potentially showing novel helimagnetic phases. This study demonstrates that computational visualization of multi-step synthetic pathways involving high pressure can accelerate the search for new metastable perovskites with rich magnetic phases.

**Introduction**

Perovskite-type iron oxides containing Fe$^{4+}$ ions have attracted attention as rare oxides showing a rich variety of helimagnetic phases. All of the perovskite-type oxides AFeO$_3$ with A = Ca, Sr, Ba show helimagnetic ordering, despite the differences in the lattice symmetry and electrical conductivity[1–3]. Notably, a cubic perovskite SrFeO$_3$ has been reported to show versatile helimagnetic orders, including multi-$q$ topological spin order[4,5]. Helimagnetic phases have also been reported in the layered-perovskite-type Fe$^{4+}$ oxides with tetragonal distortion[6–8]. Recently, Sr$_3$Fe$_2$O$_7$ has been found to show rich magnetic phases, some of which may be exotic phases with multi-$q$ spin spirals[8,9]. The origin of the helimagnetic order in Sr$_3$Fe$_2$O$_7$ and SrFeO$_3$ have been discussed in terms of magnetic frustration model[6] and the double exchange model considering strong $d$-$p$ hybridization[10], both of which are distinctive features arising from the unusually high valence Fe$^{4+}$ state[11]. Regarding the emergence of multi-$q$ spin structures in centrosymmetric systems, importance of the magnetic frustration with rotational symmetry and the itinerant nature have been pointed out[12–14].

So far, exploration of novel helimagnetic phases has been undertaken in the cubic perovskite-type iron oxides by chemical substitution to SrFeO$_3$[3,15–17]. Considering that the introduction of anisotropy is one of the simplest ways to systematically explore novel types of magnetic phases, it is interesting to synthesize perovskite-type Fe$^{4+}$ oxides (Ca,Ba)FeO$_3$ with layered ordering at the A-site. However, we have found this is rather challenging as the combination of Ca and Ba ions with the same oxidation state of +2 is not suitable for the cation ordering and the Fe$^{4+}$ ions in (Ca,Ba)FeO$_3$ tend to be even more unstable as compared with those in SrFeO$_3$. In addition, obtaining metastable perovskite-type Fe$^{4+}$ oxides is difficult in general, since it requires a multi-step synthesis via an oxygen-deficient phase. To overcome these difficulties, we performed first-principles calculations to evaluate the thermodynamic stability of putative perovskite-related compounds under ambient and high pressures[18,19]. Such calculations can be useful for the screening of target compounds and the visualization of energy landscape or synthetic pathways, potentially accelerating the exploration of new metastable materials by high-pressure synthesis[20–22].

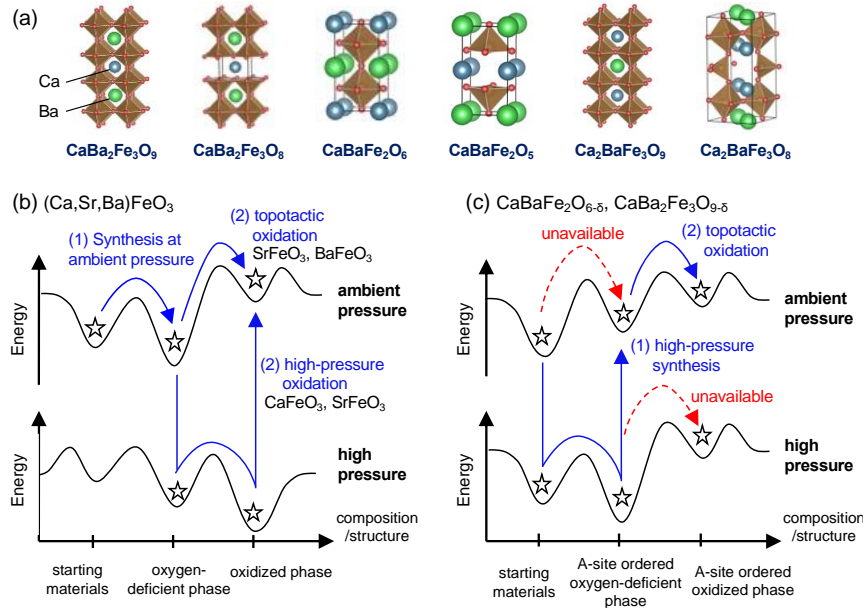

Fig. 1. (a) Crystal structure of A-site ordered perovskites $(Ca,Ba)FeO_{3-\delta}$ with and without oxygen deficiency. Schematic synthetic pathways at ambient and high pressures for (b) perovskites $(Ca,Sr,Ba)FeO_3$ and (c) A-site ordered perovskites $(Ca,Ba)FeO_3$.

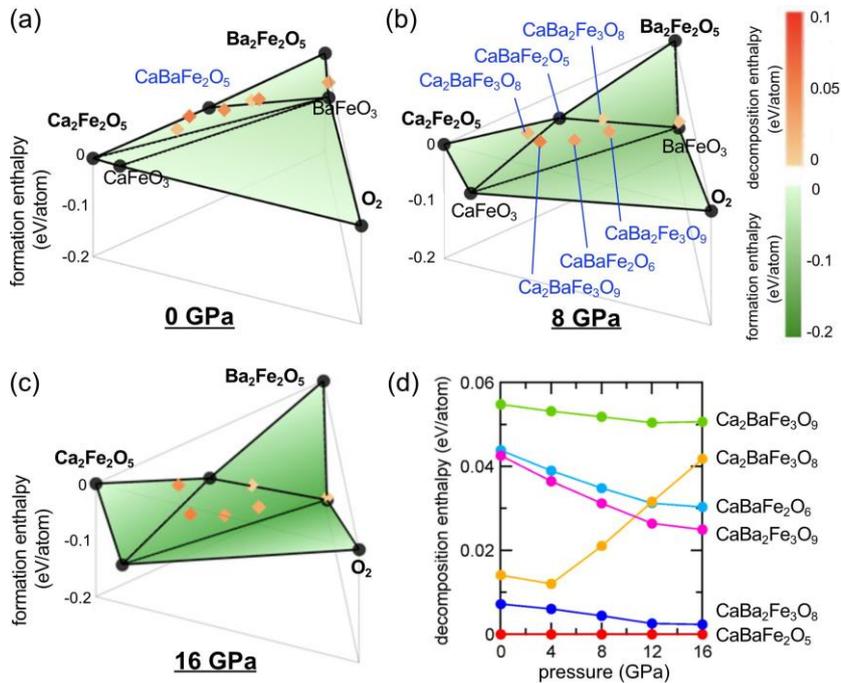

Fig. 2. Calculated convex hulls in the pseudo-ternary system of $Ca_2Fe_2O_5$–$Ba_2Fe_2O_5$–$O_2$ at (a) 0, (b) 8, and (c) 16 GPa. The black circles and colored diamond-shaped markers indicate the compounds on and above the convex hull, respectively. The convex hull is depicted as the green planes, whose brightness shows the formation enthalpy. For the compounds above the convex hull, the decomposition enthalpies, i.e., the distances from the convex hull, are indicated by the marker color. For each compound, only the result of the lowest formation enthalpy (either FM or AFM) is plotted (see also Fig. 6). (d) The pressure dependence of the decomposition enthalpies for the calculated six candidates of the new A-site ordered perovskite-type oxides.

## Results and Discussion

The scheme of the synthetic pathways for the perovskite-type iron oxides with $Fe^{4+}$ is presented in Figs. 1(b) and 1(c). $SrFeO_3$ can be obtained by high-pressure oxygen annealing or ozone annealing on the Brownmillerite-type oxygen-deficient perovskite $SrFeO_{2.5}$ phase synthesized by solid state reaction at ambient pressure, which corresponds to the blue arrows in Fig. 1(b). On the other hand, the A-site ordered perovskites $(Ca,Ba)FeO_3$ studied here could not be obtained via the same synthetic pathway (see red arrow with broken line in Fig. 1(c)).

Therefore, we performed DFT calculations to examine the thermodynamic stability of putative A-site ordered perovskite-type ferrites $(Ca,Ba)FeO_{3-\delta}$. The synthetic pathway of them was experimentally optimized, which is schematically shown in Fig. 1(c).

Figures 2(a)-(c) show the obtained convex hulls at 0, 8, and 16 GPa. The black circle markers show the compounds located at the vertex of the convex hull. Their decomposition enthalpies are equal to zero, indicating they are thermodynamically stable. The diamond-shaped markers represent the compounds floating

above the convex hull. Their decomposition enthalpies are equal to the vertical distance from the convex hull, which is visualized by the color scale. Although they are predicted to be thermodynamically unstable, one can still expect reasonably high synthesizability when the decomposition enthalpy is less than approximately 0.1 eV/atom[23].

Figure 2(d) shows the pressure dependence of the decomposition enthalpies of the assumed six candidates. Among them, the oxygen deficient compounds, $CaBaFe_2O_5$ and $CaBa_2Fe_3O_8$, have been found to have reasonably small decomposition enthalpies under high pressures, suggesting their high synthesizability by a high-pressure technique. Here, $CaBaFe_2O_5$ is predicted to be stable even under ambient pressure, its formation enthalpy decreases with the application of pressure, implying that it tends to be stabilized at high pressures. While $CaBa_2Fe_3O_8$ has a small but finite positive decomposition enthalpy over the entire pressure range, the decrease in the decomposition enthalpy with applying pressure suggests that high-pressure synthesis can serve as an effective approach for its synthesis as well as $CaBaFe_2O_5$. Contrary to these two oxygen-deficient perovskites, $Ca_2BaFe_3O_8$ was found to become unstable upon the application of pressure above 4 GPa. These results are consistent with the fact that the oxygen-deficient A-site-ordered perovskites $CaBaFe_2O_5$ and $CaBa_2Fe_3O_8$, were successfully obtained by high pressure synthesis (see the blue arrow in Fig. 1(c)), whereas $Ca_2BaFe_3O_8$ was not. On the other hand, the fully-oxidized phases $CaBaFe_2O_6$ and $Ca_2BaFe_3O_9$, are less stable at high pressures as compared with those of the oxygen-deficient counterparts, which is also consistent with our experimental failure of high pressure oxygen annealing for the oxygen-deficient phases. Therefore, we adopted topotactic oxidation at low temperatures for their corresponding oxygen-deficient phases as mentioned below.

On the basis of the DFT-based calculations of the thermodynamic stability at selected pressures, we have succeeded in obtaining the two oxygen-deficient phases, $CaBaFe_2O_5$ and $CaBa_2Fe_3O_8$. Note that $Ca_2BaFe_3O_8$ was unavailable even by the high-pressure synthesis, as suggested by the DFT calculations shown in Fig. 2(d), which is described by a red arrow with broken line in Fig. 1(c). Figs. 3(a-b) show the SXRD patterns and the results of Rietveld refinement. The refined structural parameters with the weighted profile R factor ($R_{wp}$) and the goodness of fit ($S$) are summarized in Table S1 in the supplemental information. In the weighing process of the starting materials for synthesis, the ratio of Ca:Ba was slightly adjusted to obtain $Ca(Ba_{0.9}Ca_{0.1})_2Fe_3O_8$ instead of $CaBa_2Fe_3O_8$, in order to avoid the formation of impurities. As a similar compositional deviation in A-site ordered perovskites, $Sr_{3.12}Er_{0.88}Co_4O_{10.5}$, which deviates from the ideal ordered composition $Sr_3ErCo_4O_{10.5}$, has been reported[24]. The reflection at $2\theta \sim 5.2°$ for $CaBaFe_2O_5$ and that at $2\theta \sim 3.4°$ for $Ca(Ba_{0.9}Ca_{0.1})_2Fe_3O_8$ can be indexed as 001 with respect to their tetragonal unit cell. These reflections can be regarded as the superlattice peaks corresponding to the doubling (tripling) of the cubic cell along the $c$-axis due to the formation of two-fold (three-fold) A-site ordering. The deviation of the refined A-site occupancy from the ideal occupancy assuming complete A-site ordering are 2.5 % for $CaBaFe_2O_5$ and 0.3 % for $Ca(Ba_{0.9}Ca_{0.1})_2Fe_3O_8$, suggesting the successful formation of almost complete A-site ordering. Figs. 3(c-d) show the results of Rietveld refinement after the ozone oxidation. The refined structural parameters with $R_{wp}$ and $S$ are summarized in Table S2 and S3 in the supplemental information. For $CaBaFe_2O_{6-\delta}$, the coexistence of two phases is suggested from the splitting of the main peak (see also Fig. S1). The main phase with volume fraction of 86.5 % is the partially oxidized phase, where the formal valence for Fe is estimated to be +3.6 from the refined oxygen vacancy ($\delta \sim 0.4$). The secondary phase represents the oxygen-deficient phase, which remains unoxidized. The SXRD patterns for $Ca(Ba_{0.9}Ca_{0.1})_2Fe_3O_{9-\delta}$ were fitted well by a single partially-oxidized phase, of which formal valence of Fe is estimated to be +3.6 ($\delta \sim 0.6$). The relatively large $R_{wp}$ and $S$ for $CaBaFe_2O_{6-\delta}$ reflects a presence of unknown impurity.

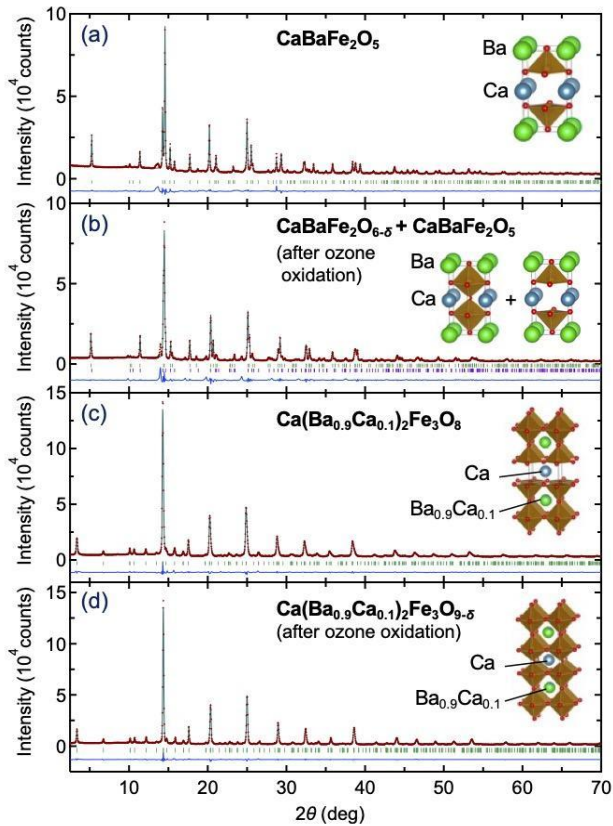

Fig. 3. The Rietveld refinements of the synchrotron XRD patterns for (a) $CaBaFe_2O_5$, (b) $CaBaFe_2O_{6-\delta}$ + $CaBaFe_2O_5$, (c) $Ca(Ba_{0.9}Ca_{0.1})_2Fe_3O_8$, and (d) $Ca(Ba_{0.9}Ca_{0.1})_2Fe_3O_{9-\delta}$. The red dots, green lines, blue lines, and vertical tick marks represent observed patterns, calculated patterns, differences between observed and calculated intensities, and Bragg reflection points, respectively. The insets show the crystal structures visualized by VESTA software[25]. The green and purple tick marks in (c) correspond to the partially-oxidized phase ($CaBaFe_2O_{6-\delta}$) and oxygen-deficient phase ($CaBaFe_2O_5$), respectively.

Figure 4 summarizes the obtained structural parameters for the oxygen-deficient phase and the sample after ozone oxidation. The decrease in the lattice constant $a$ is presumed to correspond to a decrease in the Fe-O bond length with increasing valence of the Fe ions. Although the sample is not fully oxidized to $Fe^{4+}$ state after ozone oxidation, partial oxidation was achieved as suggested by the decrease in $\delta$ and the increase in the bond valence sums (BVS) of Fe ions. Fig. 4(i) shows the BVS of the $AFeO_3$ as a function of the average ionic radius of the A-site cation $<r_A>$. Note that the downward deviation of the BVS of $Ca_{1-x}Sr_xFeO_3$ and $Sr_{1-x}Ba_xFeO_3$ from the nominal valence of Fe ions at large $<r_A>$ reflects the lattice expansion due to the increase in the average size of the A-site ions[2,3]. The BVS for the A-site ordered phases increase by ozone oxidation, but it is not as high as that of the fully oxidized perovskites with $Fe^{4+}$ such as $SrFeO_3$ and $BaFeO_3$. To further characterize the valence of Fe ions in the oxidized phases, we conducted thermogravimetric (TG) measurements for the oxidized phases under $N_2$ flow as shown in Fig. S3 in the supplemental information. Here we note that the main phase after the TG measurements is A-site-disordered oxygen-deficient perovskite phases, and that in the previous report, $SrFeO_{3-d}$ becomes an oxygen-deficient perovskite $SrFeO_{2.5}$ with $Fe^{3+}$ ions when heated to 1000 °C under $N_2$ atmosphere[26]. Thus, from the TG curves with assuming that the valence of Fe ions becomes +3 after the TG measurements, the $\delta$ values in $CaBaFe_2O_{6-\delta}$ and $Ca(Ba_{0.9}Ca_{0.1})_2Fe_3O_{9-\delta}$ are estimated to be 0.576 and 0.257, giving rise to the nominal Fe valence as +3.42 and +3.83, respectively.

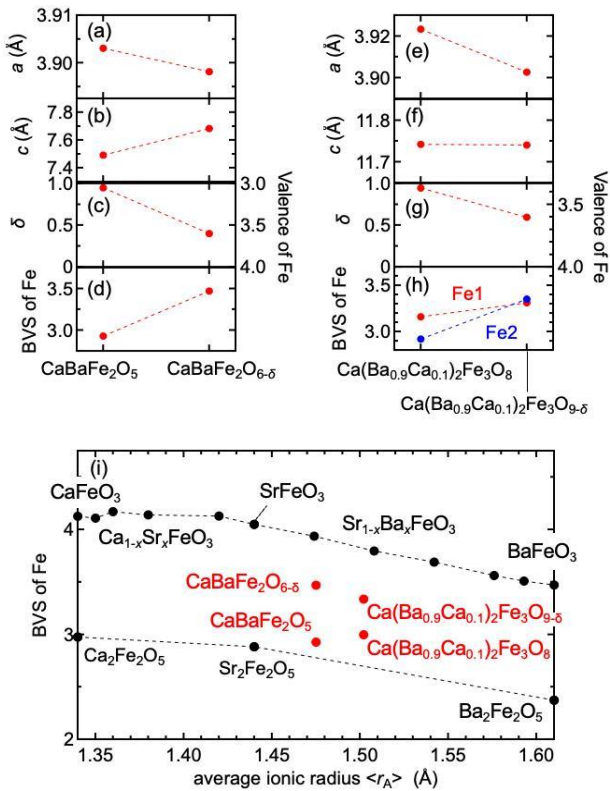

Fig. 4. (a-h) The lattice constant $a$, $c$, the oxygen deficiency $\delta$, and the bond valence sums (BVS) of Fe ions obtained from the Rietveld refinement for $CaBaFe_2O_{6-\delta}$ and $Ca(Ba_{0.9}Ca_{0.1})_2Fe_3O_{9-\delta}$ before and after ozone oxidation. (i) The BVS of Fe ions in $(Ca,Ba)FeO_{3-\delta}$ as a function of the average ionic radius of the A-site cation together with the relevant A-site disordered perovskites; $Ca_{1-x}Sr_xFeO_3$[2] and $Sr_{1-x}Ba_xFeO_3$[3].

Figure 5 shows the result of magnetic measurements. As for the oxygen-deficient perovskite $CaBaFe_2O_5$ and the oxidized perovskite $CaBaFe_2O_{6-\delta}$ with the two-fold A-site ordering, Curie-Weiss like behavior is absent even near room temperature. In addition, the magnetization curve at 300 K is nonlinear, indicating the occurrence of a canted antiferromagnetic transition above 300 K. There also exists another transition near 80 K, probably corresponding to spin reorientation. Comparing the magnetization curves at 2 K between $CaBaFe_2O_5$ and $CaBaFe_2O_{6-\delta}$, the magnetization of the latter at 7 T is twice as large as that of the former, which suggests the change in the ground state from AFM to a more FM-like phase. Although we should consider the effect of a small amount of impurity phases, the emergence of the FM-like phase is also consistent with our DFT calculations, as discussed later in Fig. 6. $Ca(Ba_{0.9}Ca_{0.1})_2Fe_3O_8$ and $Ca(Ba_{0.9}Ca_{0.1})_2Fe_3O_{9-\delta}$ with three-fold A-site ordering also show non-Curie-Weiss behavior near 300 K, implying the canted antiferromagnetism as well as the two-fold A-site ordered phases. We note that the oxidized perovskite $Ca(Ba_{0.9}Ca_{0.1})_2Fe_3O_{9-\delta}$ shows an additional transition near 50 K, which is absent in $Ca(Ba_{0.9}Ca_{0.1})_2Fe_3O_8$. The anomaly at 180 K is likely attributed to $CaFe_2O_4$, slightly mixed as an impurity[27], while it is invisible in the SXRD pattern.

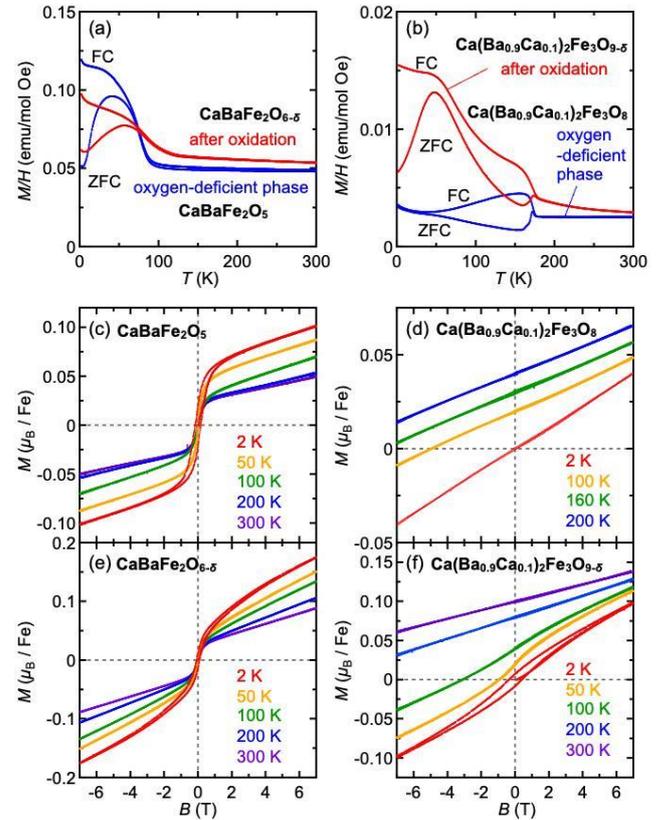

Fig. 5. The temperature dependence of magnetic susceptibility of (a) $CaBaFe_2O_{6-\delta}$ and (b) $Ca(Ba_{0.9}Ca_{0.1})_2Fe_3O_{9-\delta}$ and the corresponding oxygen-deficient phases. The magnetic field dependence of magnetization for (c) $CaBaFe_2O_5$, (d) $Ca(Ba_{0.9}Ca_{0.1})_2Fe_3O_8$, (e) $CaBaFe_2O_{6-\delta}$, (f) $Ca(Ba_{0.9}Ca_{0.1})_2Fe_3O_{9-\delta}$.

In order to discuss the magnetic ground states of the newly obtained A-site ordered perovskites, we performed DFT calculations to estimate the decomposition enthalpy assuming FM and AFM ordering as shown in Fig. 6. For the oxygen-deficient compounds such as $CaBaFe_2O_5$ with $Fe^{3+}$ and $CaBa_2Fe_3O_8$ with $Fe^{3.33+}$, an AFM state is predicted to be more stable than FM ordering. This is consistent with the fact that most iron oxides with $Fe^{3+}$ are AFM insulators, as exemplified by $LaFeO_3$. On the other hand, for the oxidized compounds with $Fe^{4+}$ such as $CaFeO_3$, $CaBa_2Fe_3O_9$, and $CaBaFe_2O_6$, a FM state has smaller decomposition enthalpy than a AFM state. Since iron oxides with $Fe^{4+}$ such as $CaFeO_3$ and $BaFeO_3$ have been reported to exhibit helimagnetic ordering, the magnetic ground states of $CaBa_2Fe_3O_9$ and $CaBaFe_2O_6$ can be helimagnetic as well. As shown in Fig.5(b), $Ca(Ba_{0.9}Ca_{0.1})_2Fe_3O_{9-\delta}$ was found to show antiferromagnetic ordering at 50 K, which is absent in the oxygen-deficient phase of

Ca(Ba$_{0.9}$Ca$_{0.1}$)$_2$Fe$_3$O$_8$. Thus, it is reasonable to presume that the magnetic anomaly at 50 K in Ca(Ba$_{0.9}$Ca$_{0.1}$)$_2$Fe$_3$O$_{9-\delta}$ corresponds to the emergence of a novel magnetic, potentially helimagnetic phase. As for CaBaFe$_2$O$_{6-\delta}$, since a magnetic anomaly around 80 K is found not only in CaBaFe$_2$O$_{6-\delta}$ but also in CaBaFe$_2$O$_5$, it is difficult to discuss the emergence of helimagnetic ordering therein. In a layered square lattice with staggered DM interactions, emergence of a 2-$q$ chiral stripe phase is predicted by simulated annealing[28]. Such a multi-$q$ phase is expected to appear in these highly oxidized perovskite-type ferrites with tetragonal distortion.

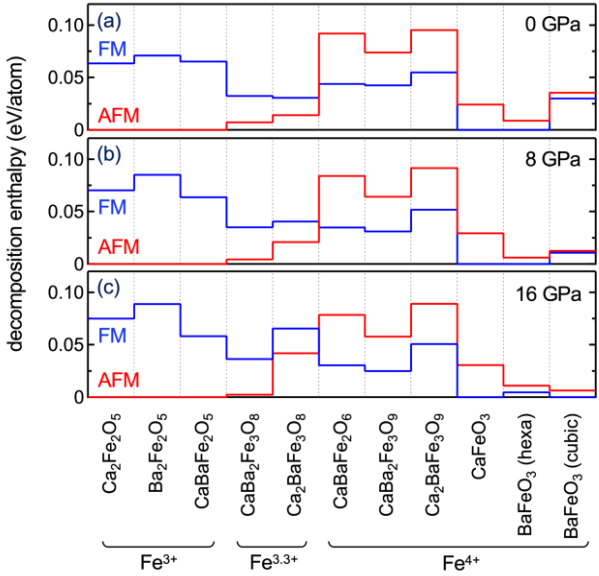

Fig. 6. The calculated decomposition enthalpies with assuming ferromagnetic (FM) and antiferromagnetic (AFM) order under (a) 0 GPa, (b) 8 GPa, and (c) 16 GPa.

Finally, we discuss the mechanism responsible for the stabilization of the A-site cation ordering in the present perovskites (Ca,Ba)FeO$_{3-\delta}$. The A-site ordering is typically formed when the valence state of the two A-site cations are different from each other mostly by 1 or 2. However, in CaBaFe$_2$O$_5$ and Ca(Ba$_{0.9}$Ca$_{0.1}$)$_2$Fe$_3$O$_8$, the uncommon A-site ordering by isovalent Ca$^{2+}$ and Ba$^{2+}$ is achieved. It has been reported that the cation ordering in perovskites is often coupled with oxygen vacancy ordering[29–31], which can be an important factor to stabilize the isovalent A-site ordering in the present systems. Due to the oxygen vacancy ordering along the $c$ axis, A-sites are split into 12-coordinated sites and 8-coordinated sites. While the former accommodate the larger Ba$^{2+}$ ions, the latter host the smaller Ca$^{2+}$ ions, resulting in the cooperative stabilization between the oxygen vacancy ordering and the A-site ordering. This effect can be enhanced by the application of pressure, as suggested by our DFT calculations of structural optimization for three different types of A-site ordering (see Fig. S2 in the supporting information). This seems to be consistent with the fact that Ca(Ba$_{0.9}$Ca$_{0.1}$)$_2$Fe$_3$O$_8$ and CaBaFe$_2$O$_5$ could only be synthesized under high pressure.

In summary, we have successfully synthesized two types of new A-site ordered oxygen-deficient perovskites, CaBaFe$_2$O$_5$ and Ca(Ba$_{0.9}$Ca$_{0.1}$)$_2$Fe$_3$O$_8$, with the aid of enthalpy calculations for selected pressures. We also tried topotactic oxidation to these oxygen-deficient phases by ozone annealing at low temperatures. Although the samples were not fully oxidized to Fe$^{4+}$ state, partial oxidation was achieved. In Ca(Ba$_{0.9}$Ca$_{0.1}$)$_2$Fe$_3$O$_{9-\delta}$ after oxidation, the emergence of a possibly helimagnetic phase is suggested. The present work demonstrates that DFT calculations can be effectively used for the screening of metastable perovskite-related oxides and visualization of synthetic pathways involving high-pressure synthesis.

**Experimental Methods**

As shown in Fig. 1(a), we assumed six types of perovskites with and without oxygen vacancies as the potential candidates for new A-site ordered perovskite-type ferrites. The candidates contain three oxidized phases (CaBa$_2$Fe$_3$O$_9$, CaBaFe$_2$O$_6$, and Ca$_2$BaFe$_3$O$_9$), and their corresponding oxygen-deficient phases (CaBa$_2$Fe$_3$O$_8$, CaBaFe$_2$O$_5$, and Ca$_2$BaFe$_3$O$_8$). The initial structures for the oxygen-deficient phases were generated by substituting elements from previously reported compounds: YBa$_2$Fe$_3$O$_8$[32] for CaBa$_2$Fe$_3$O$_8$, YBaFe$_2$O$_5$[33] for CaBaFe$_2$O$_5$, and Ca$_2$LaFe$_3$O$_8$[34,35] for Ca$_2$BaFe$_3$O$_8$. DFT-based structural optimization calculations were performed using Quantum Espresso package [36,37] under 0, 4, 8, 12, and 16 GPa. Subsequently, formation enthalpies for each compound were computed in the pseudo-ternary system of Ba$_2$Fe$_2$O$_5$ – Ca$_2$Fe$_2$O$_5$ – O$_2$. Finally, decomposition enthalpies for each compound were calculated by generating a convex hull using the PhaseDiagram module[38] implemented in Pymatgen package[39].

The computational details of the DFT-based structural optimization calculations were as follows. We used the PBE-type generalized gradient approximation (GGA) as an exchange–correlation functional and the pseudopotentials from the Standard Solid State Pseudopotentials library (SSSP Precision)[40,41]. Cutoff energies for the wavefunction and charge density were set to 90.0 Ry and 1080.0 Ry. The Hubbard $U$ parameter for the Fe atom was set to 4 eV. We performed the calculations assuming ferromagnetic (FM) and G-type antiferromagnetic (AFM) order for each compound, and used the one with lower enthalpy for the generation of the convex hull.

Based on the above calculations of thermodynamic stability, we tried high-pressure synthesis for the two oxygen-deficient A-site ordered perovskites CaBaFe$_2$O$_5$ and CaBa$_2$Fe$_3$O$_8$, which were predicted to be stabilized under high pressure. As for CaBa$_2$Fe$_3$O$_8$, the composition was slightly modified to Ca(Ba$_{0.9}$Ca$_{0.1}$)$_2$Fe$_3$O$_8$ at the weighing procedure to suppress the formation of impurities with hexagonal perovskite-type structure. First, starting materials (BaCO$_3$, CaCO$_3$, and Fe$_2$O$_3$) were stoichiometrically mixed and fired in air at 1100 °C for 12 h. The obtained precursor was the mixture of Ca$_2$Fe$_2$O$_5$, BaFe$_2$O$_4$, BaO, and (Ca,Ba)FeO$_{3-\delta}$ with A-site disordered cubic perovskite-type structure. Then, the precursor was treated at high pressure and high temperature using cubic-anvil type high-pressure apparatus. In the case of Ca(Ba$_{0.9}$Ca$_{0.1}$)$_2$Fe$_3$O$_8$, the precursor was sealed in a gold capsule and heated at 8 GPa and 1200 °C for 30 min. For CaBaFe$_2$O$_5$, the precursor, sandwiched by a reducing agent of Ti powder, was sealed in a platinum capsule, and heated at 8 GPa and 1300 °C for 30 min. For both samples after the high-pressure and high-temperature synthesis, the temperature was gradually reduced to 450 °C over a period of 6 h, followed by the quench to room temperature. Topotactic oxidation of the obtained oxygen-deficient samples was performed using an ozonizer (EcoDesign LOG-LC15G). The sample was pulverized and exposed to ozone of ~ 10 vol % at 200 °C for 3 h. The ozone was generated by silent discharge of oxygen, and flowed at 0.5 L/min.

The synchrotron powder x-ray diffraction (SXRD) measurement was performed at BL-8A, KEK, Japan. The wavelength of the x-ray was 0.68592-0.69125 Å, calibrated for each compound using $CeO_2$ standard. The Rietveld analysis of the SXRD pattern was performed using RIETAN-FP[42]. Thermogravimetric measurements were performed under $N_2$ atmosphere using an instrument of NETZSCH STA 2500. The measurement of magnetic properties was performed by Magnetic Property Measurement System (MPMS), Quantum Design.

**Supporting Information**

Tables of crystallographic data; additional results on Rietveld analysis for $CaBaFe_2O_{6-\delta}$ and enthalpy calculations for $CaBaFe_2O_5$; thermogravimetric analysis for $CaBaFe_2O_{6-\delta}$ and $Ca(Ba_{0.9}Ca_{0.1})_2Fe_3O_{9-\delta}$.

Corresponding Author
* ishiwata.shintaro.es@osaka-u.ac.jp

**Acknowledgments**

This work was partly supported by JSPS, KAKENHI (Grants No. 22H00343, 22J13408, 23H04871, and 24K00570), the Murata Science Foundation and Asahi Glass Foundation. The synchrotron powder XRD measurements were performed with the approvals of the Photon Factory Program Advisory Committee (Proposal No. 2021S2-004).